# Collective eigenstates of emission in an *N*-entity heterostructure and the evaluation of its Green tensors and self-energy components

Murugesan Venkatapathi*

*Computational Photonics Laboratory, SERC, Indian Institute of Science, Bangalore, 560012*

*murugesh@serc.iisc.ernet.in

ABSTRACT:

Our understanding of emission from a collection of emitters strongly interacting among them and also with other polarizable matter in proximity has been approximated by independent emission from the emitters. This is primarily due to our inability to evaluate the self-energy matrices and the collective eigenstates of emitters in heterogeneous ensembles. A method to evaluate the self-energy matrices that is not limited by the geometry and the material composition is presented here to understand and exploit such collective excitations. Numerical evaluations using this method are used to highlight the significant differences between independent and the collective modes of emission in heterostructures. A set of *n* emitters driving each other and *m* other polarizable entities, where *N=m+n*, is used to represent the coupled system of a generalized geometry in a volume integral approach. Closed form relations between the Green tensors of entity pairs in free space and their correspondents in a heterostructure are derived concisely. This is made possible for general geometries because the *global* matrices consisting of all free-space Green dyads are subject to conservation laws. The self-energy matrix of the emitters can then be assembled using the evaluated Green tensors of the heterostructure, but a decomposition of its components into their radiative and non-radiative decay contributions is non-trivial. This is accomplished using matrix decomposition identities applied to the global matrices containing all free-space dyads. The relations to compute the observables of the eigenstates (such as quantum efficiency, power/energy of emission, radiative and non-radiative decay rates) are presented. We conclude with a note on extension of this method to collective excitations that also include strong interactions with a surface in the near-field.

1. INTRODUCTION:

The interaction of atoms, fluorescing molecules and quantum dots with metal particles/surfaces have been extensively studied for their effect on the decay rates and radiated power of spontaneous emission.[1-8] But such understanding of the effects of surrounding matter on spontaneous emission has been limited to an *independent* emitter interacting with a particle,[4] surface,[3] or a set of spherical particles;[8] generalized as Purcell effect. Recently, this was extended by a theoretical study to a set of many emitters interacting collectively with a single metal particle at the center.[9,10] Strongly interacting emitters present collective excitations mediated by virtual photons, and in the presence of other matter can include other virtual excitations as well; plasmons of a metal particle for example. We should expect that when the interactions between a single emitter and a body are not strong relative to the possible multilateral modes, the collective effects can manifest robustly to be observed and exploited. The impediment in understanding collective heterogeneous systems has been the difficulty of evaluating the self-energy components that include all the possible virtual interactions among the coupled emitters. This requires the evaluation of Green tensors coupling emitters *in the presence of* other polarizable matter in proximity.

The diagonal terms of the self-energy matrices of such a heterogeneous system represent the independent emitters interacting with other bodies; they are typically computed using just the self-field of a single emitter due to the other polarizable bodies. Whereas, the off-diagonal terms of the self-energy matrix include the interaction among any two emitters directly and through other matter in proximity, and here we require these Green tensors explicitly. Note that analytical expressions for Green tensors are difficult in general; [11] and hence a numerical evaluation of Green tensors is a preferred approach for complex geometries. Mode decomposition and eigen function based Green tensor representation is suitable for regular geometries, and that is seldom a possibility in realistic nanostructures. We start with a set of *n* Lorentz dipole oscillators which when coupled can be used to elucidate the conventional Dicke effect of collective emission and the radiative coupling among them. In addition, here we have *m* other polarizable entities which can be sub-volumes of a larger body in a volume integral approach. Thus this method is general and precise for arbitrary shapes, properties and a number of emitters/bodies, even in the limits where long wavelength approximations are not valid. It is shown that a problem posed by dipole sources and null field conditions at polarizable volumes can be used to evaluate the modified Green tensors coupling the *n* emitters in presence of the *m* other polarizable entities. We use matrix algebra to derive closed form relations between these modified Green tensors and the set of Green dyads coupling pairs of the *N* entities in free space (or a homogeneous background). The free-space Green dyads of dipolar entities can be trivially derived by the dyadic operations on scalar Green functions of point dipoles. Use of free space dyads of multipoles to avoid a fine discretization of a body into dipoles is also possible for very large problems. The modified Green tensors coupling the emitters are then used to assemble their self-energy matrix and evaluate their eigenstates. Determining the characteristics of emission from the ensemble involves the additional exercise of the decomposing the imaginary part of the self-energy matrices into their radiative and non-radiative parts. This allows us to evaluate the quantum efficiency, decay rates and the radiated energy/power in the ensemble, given the quantum efficiency and non-radiative relaxation of an isolated emitter. The microscopic aspects in dipole representations of certain types of emitters, [12, 13] and an explicit quantization of the emitter/bodies, [14-17] are not the subject of this work, and these may be introduced in this method without loss of generality for any special cases.

Finally, we conclude with a few numerical results highlighting that self-energy matrices of such interacting systems can reveal collective effects hitherto unexplained; some of our recent experimental observations and their analysis will be described elsewhere. Additions to this method that can include strong interactions of all the *N* entities with a surface are presented as an addendum.

2. METHODS:

Consider the independent emitters as oscillating dipole moments $p_j \mathbf{e}_j$ with a resonant frequency $\omega_o$. We assume that non-radiative relaxations at a decay rate $\Gamma_o^{nr}$ result in uncorrelated emission events at a decay rate $\Gamma_o^r$ in these individual emitters, and these events are much slower than the oscillations (i.e.) $\Gamma_o^r \ll \omega_o$, and we use classical Lorentz oscillators to represent the excitations. This section has three parts; first we introduce self-energy matrices of a collection of coupled Lorentz oscillators and their significance in studying collective behavior in an emission process. Then we extend this approach to evaluation of the self-energy matrices for cases where the emitters also interact strongly with other matter. This primarily involves evaluating the Green tensors coupling the emitters in the presence of other polarizable matter in a generalized geometry. Finally we present a technique to decompose the radiative and non-radiative contributions to each self-energy component of the self-energy matrix representing a heterogeneous mixture of emitters with other polarizable bodies. This is essential because the radiative and non-radiative decay matrices determine the observables and measurements of the collective emission.

2.1. *Self-energy matrix of coupled Lorentz oscillators*:

The emission from *n* dipole oscillators interacting with each other is well represented by a coupled system with *n* eigenstates, some of them super-radiant, and this collective phenomenon of emitters interacting directly is known as the conventional Dicke effect.[18-21] Let $p_j \mathbf{e}_j$ be the dipole moments of the interacting Lorentz oscillators at cooperative phases $\phi_j$; and *q*, *m* be the magnitude and mass of the charge in the dipoles. We derive the shifts in energy $\Delta$ and the modified decay rates $\Gamma$ of the interacting ensemble as additional components of self-energy of these emitters immersed in their common field. The mechanical force of a dipole $j$ and the force of collective driving fields from the other Lorentz dipoles have to be balanced resulting in Eq. (1); where derivatives of time are represented by corresponding number of dots in the superscript.

$$[\ddot{p}_j + \Gamma_o^r \dot{p}_j + \omega_o^2 p_j] \cdot e^{i\phi_j} - \frac{q^2}{m} \sum_{k \neq j} \mathbf{e}_j \cdot \mathbf{G}_o(\mathbf{r}_j, \mathbf{r}_k) \cdot \mathbf{e}_k p_k e^{i\phi_k} = 0 \qquad (1)$$

In the above, the interaction between dipoles are described in terms of the free-space Green dyads $\mathbf{G}_o$. When $\mathbf{G}_o(\mathbf{r}, \mathbf{r}_j; \omega) \cdot \mathbf{e}_j p_j e^{i\phi_j} = \mathbf{E}(\mathbf{r}, \omega)$, the harmonic component of electric field at $\mathbf{r}$ due to an isolated dipole $p_j \mathbf{e}_j$ at $\mathbf{r}_j$. The above can be rewritten using the Fourier expansion into the following.

$$(\omega_o^2 - \omega^2 + i\Gamma_o^r \omega) \cdot p_j e^{i\phi_j} - \sum_{k \neq j} \frac{q^2}{m} \mathbf{e}_j \cdot \mathbf{G}_o(\mathbf{r}_j, \mathbf{r}_k; \omega) \cdot \mathbf{e}_k p_k e^{i\phi_k} = 0 \qquad (2)$$

For all $\omega \approx \omega_o$, it simplifies to

$$(\omega_o - \omega + \frac{i\Gamma_o^r}{2}) \cdot p_j e^{i\phi_j} - \sum_{k \neq j} \frac{q^2}{2m\omega} \mathbf{e}_j \cdot \mathbf{G}_o(\mathbf{r}_j, \mathbf{r}_k; \omega) \cdot \mathbf{e}_k p_k e^{i\phi_k} = 0 \qquad (3)$$

By substitution of $\Delta_{jk}$ and $i\Gamma_{jk}/2$ for the real and imaginary parts of the dipole cross-interaction terms ($j \neq k$) in Eq. (3), we get Eq. (4) using a phase conjugation. The system of equations reduces to a form where the *energy shifts* and *modified decay rates* of a dipole due to its interaction with other dipoles can be readily interpreted.

$$\sum_{k}[(\omega_o - \omega - \frac{i\Gamma_0^r}{2})\delta_{jk} + \Delta_{jk} - \frac{i\Gamma_{jk}}{2}] \cdot p_k e^{-i\phi_k} = 0 \qquad (4)$$

It is convenient to represent the above coupled system of equations by the matrix eigenvalue problem in Eq. (5). The unknown cooperative phases $\phi_j$ for the known oscillating dipole moments $p_j$ are given by the collective eigenstates of the oscillators that are solutions for Eq. (4). These eigenstates are represented by the eigenvalues and eigenvectors of the self-energy matrix $\Sigma$ defined in Eq. (5), after integration around $\omega_o$. The real part of its eigenvalues represents the energy shifts of a collective eigenstate, and corresponding decay rates are given by its imaginary part. Thus they represent a system of coupled equations that are fully satisfied by $(-\lambda I + \Sigma)V = 0$ where $\lambda$ are the eigenvalues and $V$ is the set of eigenvectors representing the eigenstates. A set of oscillators with known starting phases $\phi_j$ driving each other are equally well represented by Eq. (1), and the resulting dipole moment amplitudes $p_j$ can be determined using the eigenvectors of this self-energy matrix. Further, the excitation in the dipole oscillators can be given the total energy of a quantum of radiation.

Let $\Sigma_{jk}(\omega) = \Delta_{jk} - \dfrac{i\Gamma_{jk}}{2} = \dfrac{-q^2}{2m\omega}\mathbf{e}_j \cdot \mathbf{G}_o(\mathbf{r}_j, \mathbf{r}_k; \omega) \cdot \mathbf{e}_k - \delta_{jk}\dfrac{i\Gamma_0^r}{2}$ \qquad (5a)

then $[(\omega_o - \omega)\delta_{jk} + \Sigma_{jk}] \cdot p_k e^{-i\phi_k} = 0$ \qquad (5b)

also where

$\Gamma_o^r = \dfrac{2kq^2\omega_o}{3mc^2} = \dfrac{2k^3 p^2}{3\varepsilon_o \hbar}$ and $\dfrac{p}{q} = (\dfrac{\hbar}{m\omega_o})^{1/2}$ \qquad (6)

The interaction with another emitter results in energy shifts given by the *real* parts of the off-diagonal terms of the self-energy matrix, and the corresponding change in decay rates are given by the *imaginary* parts. The diagonal terms of the self-energy matrix represents the case of the independent or the uncoupled emitter and contains only its radiative decay rate. Energy shifts (or a change in resonant frequency to $\omega \neq \omega_o$) appear even for the independent emitter when it interacts with other bodies, as seen in the next section. Energy shifts in the emission can be interpreted as an equivalent shift in the ratio of the electrical self-interaction and mass of the dipole emitters. Note that the above relations dependent on the particular definition of $\mathbf{G}_o(\mathbf{r}_j,\mathbf{r}_k;\omega)$ and any other approximations used. Specifically, the Green dyads $\mathbf{G}_o(\mathbf{r}_j,\mathbf{r}_k;\omega)$ of the point dipole oscillators in free space (or a homogeneous medium) used here in Eq. (1) are derived from the following.

$$k_o^2 \varepsilon_o \mathbf{G}_o(\mathbf{r},\mathbf{r}_j;\omega) - \nabla \times \nabla \times \mathbf{G}_o(\mathbf{r},\mathbf{r}_j;\omega) = \delta(\mathbf{r}-\mathbf{r}_j) \qquad (7)$$

From Eq. (2), a self-energy matrix of the interacting Lorentz dipoles can be calculated in more generality by relaxing the $\omega \approx \omega_o$ criterion. Eq. (2) can be rewritten in a more recognizable form as in Eq. (8) after a multiplication by $(\omega_o - \omega + i\Gamma_o^r/2)$ and division by $(\omega_o^2 - \omega^2 + i\Gamma_o^r\omega)$. The form of the self-energy matrix in Eq. (9) presents a case where the dipole moments are explicitly normalized by the Lorentz factors. Also note that this normalized form of $\Sigma(\omega)$ involves integration with $(\omega_o - \omega + i\Gamma_o^r/2) \cdot d\omega$ to evaluate $\Sigma$ (and its diagonal $-\delta_{jk}i\Gamma_o^r/2$ are included after this integration).

$$(\omega_o - \omega + \frac{i\Gamma_o^r}{2}) \cdot p_j e^{i\phi_j} - (\omega_o - \omega + \frac{i\Gamma_o^r}{2})\sum_{k \neq j}\frac{q^2}{m}\mathbf{e}_j \cdot \mathbf{G}_o(\mathbf{r}_j,\mathbf{r}_k;\omega) \cdot \mathbf{e}_k \frac{p_k e^{i\phi_k}}{(\omega_o^2 - \omega^2 + i\Gamma_o^r\omega)} = 0 \qquad (8)$$

$$\Sigma_{jk}(\omega) = \Delta_{jk} - \frac{i\Gamma_{jk}}{2} = [\frac{-q^2/m}{\omega_o^2 - \omega^2 + i\Gamma_o^r\omega}]\mathbf{e}_j \cdot \mathbf{G}_o(\mathbf{r}_j,\mathbf{r}_k;\omega) \cdot \mathbf{e}_k - \delta_{jk}\frac{i\Gamma_o^r}{2} \qquad (9a)$$

$$\Rightarrow [(\omega_o - \omega)\delta_{jk} + \Sigma_{jk}] \cdot p_k e^{-i\phi_k} = 0 \qquad (9b)$$

In the theoretical approach so far, internal non-radiative relaxations of the dipole emitters have not been included, and these can be explicitly added in the diagonal terms of $\Sigma$ just as the radiative decay rates of the independent emitter (see Eq. (23)). This involves the assumption that internal non-radiative relaxation of an emitter is independent of its radiative decay. Thus using a harmonic oscillator to represent only the single quantum of emission is the typical approach. Alternately, the Lorentz oscillators can be assigned an energy that includes both the radiative and non-radiative decay of the emission. This allows for the non-radiative coupling of two emitters in proximity. The total mechanical energy of the oscillator in this case increases by a factor $1/Q_o$ while the radiative decay includes only a quantum of energy as in Eq. (10). The method presented in this paper is amenable to both definitions of the oscillator. The latter are useful in cases when emitters are also coupled non-radiatively and where dipole-dipole interactions are sufficient to represent this non-radiative coupling.

$$\Gamma_o^r = \frac{2kq^2\omega_o}{3mc^2} = \frac{2Q_o k^3 p^2}{3\varepsilon_o \hbar} \text{ and } \frac{p}{q} = (\frac{\hbar}{Q_o m \omega_o})^{1/2} \tag{10}$$

$$\text{and } Q_o = \frac{\Gamma_o^r}{\Gamma_o^r + \Gamma_o^{nr}}, \Gamma_o^r + \Gamma_o^{nr} = \Gamma_o \tag{11}$$

Further, the above description of the collective emission process is restricted by only certain cases where the number of excitations participating in the collective emission process are not negligible compared to the density of optical states (DOS) available at $\sim \hbar\omega_o$, which is typically large. This situation is possible inside cavities and photonic crystals when specifically the emission energies are near the edges of a band gap,[22] and a strong excitation includes a sufficient density of such emitters in this emission process. Also, a set of emitters can collectively share fewer excitations resulting in emissions of a higher decay rate.[23] Such special cases of non-classical emission may require a more explicit quantization of the field, and fortunately do not include a wide variety of heterogeneous optical materials and their emission.

2.2. *Evaluation of Green tensors and self-energy components of a heterostructure:*

When emitters are neither in vacuum nor in a homogeneous dielectric medium and interact with other polarizable bodies in proximity, evaluation of the Green tensors coupling them is non-trivial. Using a volume integral approach, a heterogeneous volume can be represented by sub-volumes with corresponding permittivity. In the presence of *m* other polarizable volumes *much smaller* than the wavelength in dimensions, the Green dyads are to be determined using the following Maxwell's equation for a heterostructure in a homogeneous background medium

$$\omega^2 \overline{\varepsilon}(\omega) \cdot \mathbf{E}(\mathbf{r},\omega) - c^2 \nabla \times \nabla \times \mathbf{E}(\mathbf{r},\omega) = -i4\pi\omega \{ \sum_{j=1}^{n} \dot{p}(t) \mathbf{e}_j \delta(\mathbf{r}-\mathbf{r}_j) + \sum_{j=n+1}^{N=n+m} i\omega \overline{\alpha}_j(\omega) \cdot \delta(\mathbf{r}-\mathbf{r}_j) \mathbf{E}(\mathbf{r},\omega) \} \tag{12}$$

Here $\overline{\alpha}_j$ and $\overline{\varepsilon}$ are polarizability tensor of volume *j* and the permittivity tensor of the homogeneous background respectively. An analytical solution of this discrete problem, with or without solving for these modified Green dyads, is an intractable problem in general. But given

$$k_o^2 \overline{\varepsilon}(\omega) \cdot \mathbf{G}_o - \nabla \times \nabla \times \mathbf{G}_o = \mathbf{I} \tag{13}$$

The above can be rewritten into a problem of *N* oscillators coupled by $\mathbf{G}_o(\mathbf{r}_j,\mathbf{r}_k;\omega)$, the Green dyad in the homogeneous background medium. When this medium is isotropic (with permittivity $\varepsilon(\mathbf{r} \neq \mathbf{r}_j) = \varepsilon$), the discretized representation of the dipole moments and electric fields of this heterostructure in a Cartesian coordinate system reduces to

$$\overline{G}_o \cdot P = E \tag{14}$$

where

$$\overline{G}_o = \begin{bmatrix} \overline{G}_o^{ee}(j,k) = \mathbf{G}_o(\mathbf{r}_j,\mathbf{r}_k;\omega) & | & \overline{G}_o^{eb}(j,k) = \mathbf{G}_o(\mathbf{r}_j,\mathbf{r}_k;\omega) \\ j,k=1...n & | & j=1...n, k=1:m \\ size: 3n \times 3n & | & size: 3n \times 3m \\ ----------- & - & -------------- \\ \overline{G}_o^{be}(j,k) = \mathbf{G}_o(\mathbf{r}_j,\mathbf{r}_k;\omega) & | & \overline{G}_o^{bb}(j,k) = \mathbf{G}_o(\mathbf{r}_j,\mathbf{r}_k;\omega) - \delta_{jk}\overline{\alpha}_j^{-1}(\omega) \\ j=1...m, k=1...n & | & j,k=1...m \\ size: 3m \times 3n & | & size: 3m \times 3m \end{bmatrix},$$

$$P = \begin{bmatrix} P^e \\ 3n \times 1 \\ --- \\ P^b \\ 3m \times 1 \end{bmatrix} = \begin{bmatrix} \left[\dfrac{-q^2/m}{\omega_o^2 - \omega^2 + i\Gamma_o\omega}\right]\cdot \mathbf{p}_j^e \\ j=1...n \\ ----------- \\ \mathbf{p}_j^b \\ j=n+1...m \end{bmatrix} \text{ and } E = \begin{bmatrix} E^e \\ 3n \times 1 \\ --- \\ E^b \\ 3m \times 1 \end{bmatrix} = \begin{bmatrix} \mathbf{E}_j^e \\ j=1...n \\ --- \\ \mathbf{0} \\ j=n+1...m \end{bmatrix} \tag{15}$$

and $\mathbf{G}_o(\mathbf{r}_j,\mathbf{r}_k;\omega) = \left[\overline{I} + \dfrac{c^2 \nabla\nabla}{\omega^2}\right]g(R)$ where $g(R) = (4\pi R)^{-1} \exp\left(i\dfrac{\omega\sqrt{\varepsilon/\varepsilon_o}}{c}R\right), R = \|\mathbf{r}_j - \mathbf{r}_i\|$

$$\tag{16}$$

Writing such global matrices containing the Green dyads coupling each pair of entities gives us a concise representation, but more importantly, helps unravel the implicit relations between Green dyads required to satisfy the conservation laws. The global matrix of Green dyads $\overline{G}_o$ can be decomposed further into four parts: $\overline{G}_o^{ee}$, direct interaction between pairs of emitters; $\overline{G}_o^{bb}$, interaction among the pairs of $m$ polarizable bodies; and $\overline{G}_o^{eb}, \overline{G}_o^{be}$, representing the interaction between an emitter-body pair which are transposes of each other (as the individual Green dyads are symmetric). $\overline{G}_o^{bb}$ has its diagonal dyads constituted by $-\overline{\alpha}^{-1}$ while $\overline{G}_o^{ee}$ has zeros in diagonal dyads correspondingly for the self-interaction terms (radiation reaction and non-radiative loss) of the independent emitter. $\Gamma_o$ can be conveniently included in the self-energy matrices as in Eq. (22, 23) shown later. The polarizability tensor $\overline{\alpha}$ can be corrected for sub-volumes of a contiguous large body using lattice dispersion relations if required.[24] Such finer volume discretization of a body is needed if its dimensions are not much smaller than the wavelength of emission. This limit becomes more stringent if two distinct bodies are closely spaced; for example when distance between centers $\rightarrow 2a$, where $a$ is radius of a spherical particle. We will revisit multipolar representations of a body later in the paper. Else, polarizability of a distinct particle much smaller than the wavelength is well approximated by its dipole polarizability; in case of an isotropic material this reduces to

$$\alpha_j(\omega) = a^3 \left[ \frac{\varepsilon(\omega,\mathbf{r}_j) - \varepsilon}{\varepsilon(\omega,\mathbf{r}_j) + 2\varepsilon} \right], \text{where } a \text{ is radius of particle} \tag{17}$$

In the above problem, the polarizations of the $m$ bodies $P^b$ and the collective self-fields at the $n$ emitters $E^e$ are unknown; but these do not have to explicitly solved for. The required Green dyads can be implicitly derived by rewriting the problem in terms of the required global matrix of green dyads $\overline{G}^{ee}$, coupling pairs of the $n$ emitters in this heterostructure and resulting in these collective self-fields $E^e$. This matrix should contain the required sum of the Green dyad in background homogeneous medium and its perturbation $\mathbf{G}_h(\mathbf{r}_j,\mathbf{r}_k;\omega)$ by the heterostructure, ordered by the block indicial definition in Eq. (15).

$$\overline{G}^{ee} \cdot P^e = E^e \text{ where } \overline{G}^{ee}(j,k) = \mathbf{G}_o(\mathbf{r}_j,\mathbf{r}_k;\omega) + \mathbf{G}_h(\mathbf{r}_j,\mathbf{r}_k;\omega) \text{ for } j,k = 1...n \tag{18}$$

Using matrix block multiplications, from Eq. (15) we get the following two relations; first is the momentum conservation relation that has to be satisfied by self-fields $E^e$ of *point* emitters as an optical theorem, [25] and the second is the null-field condition for the polarizable bodies.

$$\overline{G}_o^{ee} \cdot P^e + \overline{G}_o^{eb} \cdot P^b = E^e \tag{19a}$$

$$\overline{G}_o^{be} \cdot P^e + \overline{G}_o^{bb} \cdot P^b = 0 \tag{19b}$$

Removing $P^b$ in the above relations and substituting the result in Eq. (14), we get the global matrix with the modified Green dyads between pairs of emitters in the presence of the other polarizable bodies in this heterostructure.

$$\overline{G}^{ee} = \overline{G}_o^{ee} - \overline{G}_o^{eb} \cdot [\overline{G}_o^{bb}]^{-1} \cdot \overline{G}_o^{be} \tag{20}$$

$\overline{G}^{ee}$ has ordered dyads $\mathbf{G}(\mathbf{r}_j,\mathbf{r}_k;\omega) = \mathbf{G}_o(\mathbf{r}_j,\mathbf{r}_k;\omega) + \mathbf{G}_h(\mathbf{r}_j,\mathbf{r}_k;\omega)$ for $j,k = 1...n$, and these are used instead of $\mathbf{G}_o(\mathbf{r}_j,\mathbf{r}_k;\omega)$ in Eq. (9) to assemble the self energy matrices. Note that the diagonal entries of $\overline{G}^{ee}$ in Eq. (20) are non-zero unlike its free-space correspondent $\overline{G}_o^{ee}$; this is due to additional self-interaction of emitters due to other polarizable matter. But it is important to note why this method is vastly different from computing the modified Green dyads using a sum of all possible paths of interaction between two emitters in a heterostructure; which is the most obvious evaluation given all the Green dyads of free space or a homogeneous background. Firstly, the total number of possible paths of interaction between two emitters in an *m* body heterostructure is large; $C(_l^m)$ for a subset of *l* bodies. Evaluation of the resulting vector and phase for each of these paths includes ($4 \cdot 3^3 \cdot l$) multiplication operations; ($4 \cdot 3^3$) operations in the case of multiplying any two three dimensional Green dyads in complex number arithmetic. Correspondingly, the number of arithmetic operations in evaluating the perturbations to free-space Green dyads is large $\sim \sum_{l=1}^{m} 4 \cdot 3^3 \cdot l \cdot C(_l^m) > 2^m$, an idea useless for $m >> 1$. On the other hand evaluation using Eq. (20) involves *less* than $4.3^3.m^3$ multiplication operations at most; for $m \sim 10^4$ repeated such evaluations are possible today even with a personal computing device. A physical interpretation of Eq. (21) is thus meaningful; this perturbation to the global matrix containing free space Green dyads can be rewritten as a *minimization* problem shown below. This results from interpreting the inverse of a matrix $A$ as projector into a vector space (Krylov subspace) of a minimum *monic* (implies $c_o=1$ in Eq. (21)) polynomial of matrix $A$; expanded in the powers of the matrix $\overline{G}_o^{bb}$ as below.[26]

$$\overline{G}_o^{eb} \cdot [\overline{G}_o^{bb}]^{-1} \cdot \overline{G}_o^{be} = \min_{p_m \in P_m} \left\| \overline{G}_o^{eb} \cdot p_m(\overline{G}_o^{bb}) \cdot \overline{G}_o^{be} \right\|$$
$$= \min \left\| c_o \overline{G}_o^{eb} \cdot \overline{G}_o^{be} + c_1 \overline{G}_o^{eb} \cdot (\overline{G}_o^{bb}) \cdot \overline{G}_o^{be} + c_2 \overline{G}_o^{eb} \cdot (\overline{G}_o^{bb})^2 \cdot \overline{G}_o^{be} + .... + c_m \overline{G}_o^{eb} \cdot (\overline{G}_o^{bb})^m \cdot \overline{G}_o^{be} \right\|$$

(21)

where $p_m$ represents a polynomial of degree $m$ and the full set of such possible polynomials is $P_m$. Thus $p_m(\overline{G}_o^{bb})$ includes all powers of the global matrix of Green dyads coupling the bodies, and note that $(\overline{G}_o^{bb})^l$ contains the resulting dyads of all the possible $l$ interaction paths between them. Even paths involving more than $m$ interactions are implicitly included as they are anyway linear combinations of the above paths. Hence Eq. (21) represents a *minimum of sum over all paths* of interaction between any pair of emitters, and this method is thus equivalent to a Lagrangian solution of the problem.

2.3. *Radiative and non-radiative contributions to self energy components*:

The radiative and non-radiative parts of this collective emission have to be determined for a comparison with experimental measurements of the radiative properties and decay rates. The non-radiative losses of a collective mode of emission depend on the interactions of the emitters through the other bodies in addition to their internal non-radiative relaxations. The non-radiative absorption of an isolated body significantly smaller than wavelength is well approximated by its dipolar absorption $\Im[\alpha(\mathbf{r}_j)\mathbf{E}^*(\mathbf{r}_j)\cdot\mathbf{E}(\mathbf{r}_j)]$ and depends only on the imaginary part of $\overline{\alpha}$.[27] Here, the contributions of the imaginary parts of $\overline{\alpha}$ (of *all* the interacting polarizable bodies) to the self-energies of interaction between *any* two emitters have to be decomposed. In the *m* body system, these components contain both the real and imaginary parts of polarizability, as radiative interactions among bodies can precede a non-radiative loss. The radiative contributions of an isolated small body involve the real part and magnitude of its $\overline{\alpha}$ (i.e.) only the real part of its $\overline{\alpha}^{-1}$. Thus we have two components to the imaginary part of self-energy matrix $\Gamma_{jk}$, one that involves imaginary part of $\overline{\alpha}^{-1}$ of a body as a factor in the self-energy components, and the exclusion that represent the radiative contributions from all interacting bodies. These are represented by the non-radiative decay matrix $\Gamma^{nr}_{jk}$ and the radiative decay matrix $\Gamma^{r}_{jk}$ respectively. Evaluation of the radiative or non-radiative decay of collective eigenstates requires such a decomposition of self-energy matrix, as a function of the specific geometry defined by the global matrix of Green dyads. Using matrix decomposition identities for the inverse of sum of two full-rank matrices, we can derive these non-radiative and radiative parts of the total self-energy matrix as given below. Let

$$\Delta^{r}{}_{jk} - \frac{i\Gamma^{r}{}_{jk}}{2} = [\frac{-q^2/m}{\omega_o^2 - \omega^2 + i\Gamma^{r}_o\omega}]\mathbf{e}_j \cdot \mathbf{G}^{r}(\mathbf{r}_j,\mathbf{r}_k;\omega)\cdot\mathbf{e}_k - \delta_{jk}\frac{i\Gamma^{r}_o}{2} \qquad (22)$$

$$\Delta^{nr}{}_{jk} - \frac{i\Gamma^{nr}{}_{jk}}{2} = [\frac{-q^2/m}{\omega_o^2 - \omega^2 + i\Gamma^{r}_o\omega}]\mathbf{e}_j \cdot \mathbf{G}^{nr}(\mathbf{r}_j,\mathbf{r}_k;\omega)\cdot\mathbf{e}_k - \delta_{jk}\frac{i\Gamma^{nr}_o}{2} \qquad (23)$$

where the required Green dyads are
$\mathbf{G}^{r}(\mathbf{r}_j,\mathbf{r}_k;\omega) = \overline{G}^{r}(j,k)$ and $\qquad (24)$
$\mathbf{G}^{nr}(\mathbf{r}_j,\mathbf{r}_k;\omega) = \overline{G}^{nr}(j,k)$ for $j,k=1....n$

The required radiative and non-radiative Green dyads and their corresponding global matrices in Eq. (24) can be derived using a decomposition of $\overline{G}_o^{bb}$ containing the imaginary and real part of polarizability as in Eq. (25, 26). Our objective here is to evaluate the corresponding contributions to $[\overline{G}_o^{bb}]^{-1}$ and the Green dyads coupling emitters, $\overline{G}^{ee}$ in Eq. (20). These result in the radiative and non-radiative global dyads of Eq. (24) evaluated by Eq. (28, 29). We decompose the *global* matrix of Green dyads coupling the polarizable bodies into

$$\overline{G}_o^{bb} = \overline{G}_o^{bb1} + \overline{G}_o^{bb2} \tag{25}$$

where
$$\overline{G}_o^{bb1}(j,k) = \mathbf{G}_o(\mathbf{r}_j, \mathbf{r}_k; \omega) - \delta_{jk} \Re(\overline{\alpha}_j^{-1}(\omega)) \tag{26}$$
$$\overline{G}_o^{bb2}(j,k) = -\delta_{jk} \Im(\overline{\alpha}_j^{-1}(\omega))$$

when
$$(\overline{A} + \overline{B})^{-1} = \overline{A}^{-1} - \left(\overline{I} + \overline{A}^{-1}\overline{B}\right)^{-1} \overline{A}^{-1} \overline{B} \overline{A}^{-1} \tag{27}$$

$$\overline{G}^r = \overline{G}_o^{ee} - \overline{G}_o^{eb} \cdot [\overline{G}_o^{bb1}]^{-1} \cdot \overline{G}_o^{be} \tag{28}$$

$$\overline{G}^{nr} = \overline{G}_o^{eb} \cdot \{\overline{I} + [\overline{G}_o^{bb1}]^{-1} \cdot \overline{G}_o^{bb2}\}^{-1} \cdot [\overline{G}_o^{bb1}]^{-1} \cdot \overline{G}_o^{bb2} \cdot [\overline{G}_o^{bb1}]^{-1} \cdot \overline{G}_o^{be} \tag{29}$$

$\Gamma_o^{nr}$, the non-radiative relaxation of the independent emitter can be explicitly introduced in the self energy and non-radiative decay matrices; and we assume the quantum efficiency $Q_o$ and radiative rate of an independent isolated emitter $\Gamma_o^r$ are its only known properties other than $\hbar\omega_o$. Note that diagonal entries of the self-energy matrix in Eq. (22, 23) include the explicitly added decay rates of the independent emitter as its imaginary parts, and in addition, the evaluated decay rates and the energy shifts due to the other bodies in proximity (i.e.) the Purcell effect on independent emitters.

When $\overline{G}_o^{bb2}$ is not a full rank matrix because one or more of the $m$ polarizable volumes have $\Im(\overline{\alpha}) = \overline{0}$, Eq. (27) can be replaced by other matrix decomposition identities. Similarly note that $\overline{G}_o^{bb1}$ is invertible when all the real parts of polarizability of the bodies are non-zero, and has the same requirement in such special cases otherwise. Formulae for inverting a sum of an invertible and a rank 1 matrix can be recursively used where matrix $\overline{B}$ in $(\overline{A}+\overline{B})^{-1}$ need not be invertible and of low rank.[28-30] Similarly matrix deflating techniques are also possible when the real or imaginary part of the polarizability of some of the bodies is zero, and these methods can be found elsewhere.[31] The self-energy components $\Sigma_{jk}(\omega)$ (excluding diagonal terms) in Eq. (22, 23) are numerically integrated over $(\omega_o - \omega + i\Gamma_o^r/2) \cdot d\omega$, to evaluate the total self-energy matrix and decay rates. The eigenstates $J$ and their radiative decay rate are respectively given by[10]

$$\Sigma |J\rangle = \Delta_J - \frac{i\Gamma_J}{2}|J\rangle \tag{30}$$

$$\Gamma_J^r = \langle J | \Gamma^r | J \rangle \tag{31}$$

The normalized quantum efficiency $Q_h$ and *power* of emission $P_h$ of the heterostructure are given by the sum over the corresponding values of eigenstates as below.[10]

$$Q_h = \frac{1}{n}\sum_J \frac{\Gamma_J^r}{\Gamma_J^r + \Gamma_o^{nr}} \tag{32}$$

$$P_h = \sum_J Q_J \Gamma_J^r \tag{33}$$

The total rate of decay from an ensemble of heterostructures as observed by lifetime measurements can be traced using the energy radiated by all eigenstates. In determining the power and tracing the decay as in Eq. (33, 34) note that we do not evaluate the decay of a specific initial state in a particular geometry of heterostructure. That requires expanding the prepared initial state using a weighted sum of the eigenstates, and this is not of relevance in experiments where initial states of the system are unknown and random (due to the non-radiative process that accompanies). Alternately, we can determine the eigenstates for all possible random permutations in the geometry of a heterostructure along with random orientations of its dipole emitters. Once the collective eigenstates of a particular random geometry are determined, we sum the emitted power and the decay of all its eigenstates. These quantities represent an average of these observables over the full phase space of possible initial states in the geometry, which are further summed over all random geometries. These results can be directly compared to experimental measurements involving an ensemble of such structures.

$$I(t) = \sum_J Q_J \, e^{-\Gamma_J t} \tag{34}$$

Before we present numerical results of this evaluation, we conclude this section with a note on few limiting cases and the corrections required. The limits of the dyadics of point dipoles used and the effects of discretization of a body have to be discussed. This method is general enough to include arbitrary local volume discretization of a body in *multiple* scales for problems with any special cases. One such case is when two small bodies are closely spaced and a finer discretization of the body is required to include multipole interactions; for example when distance between centers → $2a$; where $a$ is radius of a spherical particle. To include $l$-pole effects sufficiently, a finer discretization of a body in dipolar representation scales the global matrices by $l^3$ in general, and thus computation by $(l^3)^3 = l^9$. Explicit use of higher order Mie modes of a sphere, its $l$-pole polarizability $\alpha_l$ and its Green dyads in the discretized representation of geometry to be evaluated, is also possible for large spherical sub-volumes. For if more Green dyads of modes up to $l$ for any polarizable volume are explicitly included in the global matrix in Eq. (15), the matrix dimensions increase by a factor of only $l$ and this will result in an increase in computation by a factor of only $l^3$. Also, one should expect that when the distance between the surface of a body and an emitter is on the order of charge separations $d = p/q$, an evaluation using the dyads of point dipoles in Eq. (13) may be not accurate. At these small separations ~ 1 *nm*, charge screening in the body may not be complete, local inhomogeneity of $\bar{\varepsilon}$ may not be negligible and electron-hole pairs can be created. Such energy transfer mechanisms can result in coupling on the order ~$r^4$ and higher which may not be represented sufficiently by finer discretization of the body alone. However, studies show that these local deviations are significantly suppressed by both non-linear effects and quantization, resulting in a freezing of higher order classical modes.[32] These effects can result in a domination of the radiative terms on the interaction with the polarizable matter even at these close separations.[33, 34] Nevertheless, corrections to $\mathbf{G}_o(\mathbf{r}_j, \mathbf{r}_k; \omega)$ for these mechanisms of interaction may be included for emitter-body separations less than 1 *nm* in any special cases. Modifications required to $\mathbf{G}_o(\mathbf{r}_j, \mathbf{r}_k; \omega)$ in case of a strongly interacting surface in the near-field of the heterostructure are presented as an addendum in this paper.

3. EXAMPLES AND NUMERICAL RESULTS:

This section presents numerical results to highlight the significance of collective emission characteristics possible when multiple emitters interact with multiple bodies. Numerical evaluations of collective eigenstates of emission from emitter-metal nanoparticle ensembles were performed. We compare the results of the method presented here (named *NS*) with two other evaluations of the same structures: 1) *independent emitters* interacting with multiple metal particles (named *IE* for independent emitters) and 2) a set of emitters collectively interacting with many *independent metal particles* where interaction among metal particles is ignored (named *EPS*, for Extended Pustovit-Shahbazyan model explained in appendix). The objective is to highlight that while the differences with the former show the significance of collective modes of emission in such heterostructures, the differences with the latter emphasize that the collective modes of emitters are sensitive to an increase in local density of optical states (LDOS) due to interactions among many metal particles. Moreover, these results are shown to be approximated by (1) in case coupling among the emitters is weak relative to the available LDOS, and to (2) in case of interaction of the collection of emitters with one or a few metal particles.

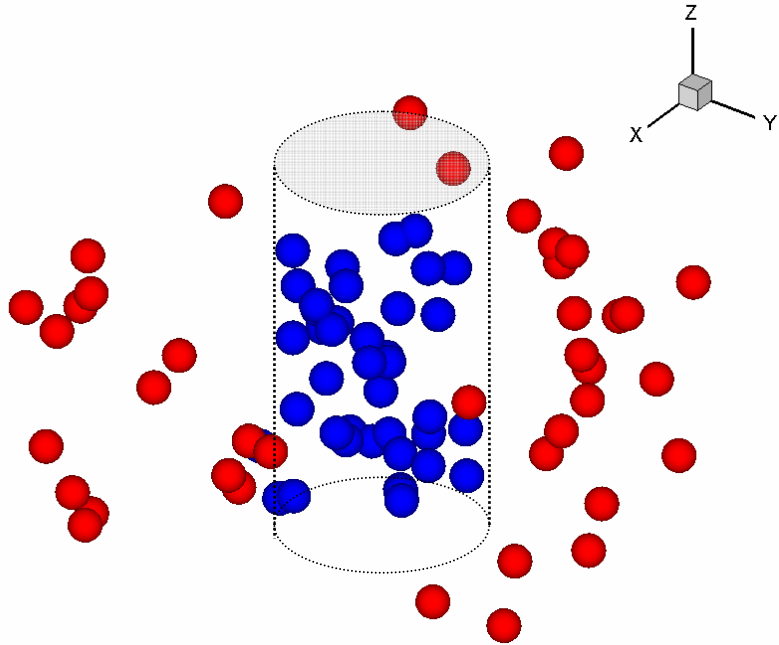

FIG. 1. A sketch of the heterostructure (not to scale); In G1 blue spheres represent the emitters and the red spheres represent metal particles, and vice versa in G2.

Consider a structure where the emitters are randomly distributed inside a cylinder of radius 20 nm and height 40 nm, while the metal particles are all randomly distributed outside this cylinder; see Figure 1. The position vectors $\mathbf{r}_p$ of the metal particles represent a random *normal* distribution in $\Re^2$ with a mean distance of 35 nm from the axis of the cylinder (and a standard deviation of 3nm).

The distribution of these particles in *z* direction (parallel to the axis of the cylinder) is a random *uniform* distribution in $\Re^1$. $\mathbf{e}_j$ are *unit* vectors of a random uniform distribution in $\Re^3$ representing the emitter polarizations. The position vectors $\mathbf{r}_e$ of the emitters represent a different random *uniform* distribution in $\Re^3$ throughout the interior space of the cylinder. Such ordered self-assembled films consisting of quantum dots and metal particles are in fact studied experimentally.[7] This geometry is named 'G1', and an inversion of such distributions where the emitters are outside the cylinder while metal particles are inside, is named 'G2'.

Using the computed collective eigenstates, the results presented include the power enhancements evaluated, relative decay rates, the quantum efficiencies and the time traces of decay using Eq. (30-34). The results produced here involve gold spheres 3 nm in radii and dipole emitters with $\hbar\omega_o = 2.21$ eV (that corresponds to a free-space wavelength of ~560 nm). Isolated emitters are assumed to have quantum efficiency of 1/6, and as shown later $Q_o$ has a almost linear effect on the quantum efficiency of the heterostructure and there is no loss of generality. The power of emission due to a continuous excitation, and the lifetimes are the typically measured variables in an experimental study of characteristics of emission. Figure 2a shows the power enhancement in the G1 heterostructure relative to an equal number of isolated emitters; there are two clear regimes apparent in this plot. These results are ensemble averages over *150* random permutations of the emitters and metal particles.

Firstly, in the limit of a few metal particles ($m \leq 10$), the full evaluation (NS) matches with the EPS evaluation of *m* independent metal particles interacting with the collection of emitters. Here the collective interaction among emitters results in a significant increase in the power enhancement, and this is not revealed in the independent emitter (IE) characteristics. As the number of particles increases further ($m \geq 60$), strong interactions among metal particles can result in a large increase in local density of optical states (LDOS); this breaks the accuracy of the EPS evaluation. This regime is dominated by the LDOS available to the emitter, and the interactions between emitters seem to play a weaker role compared to the interactions between the metal particles.

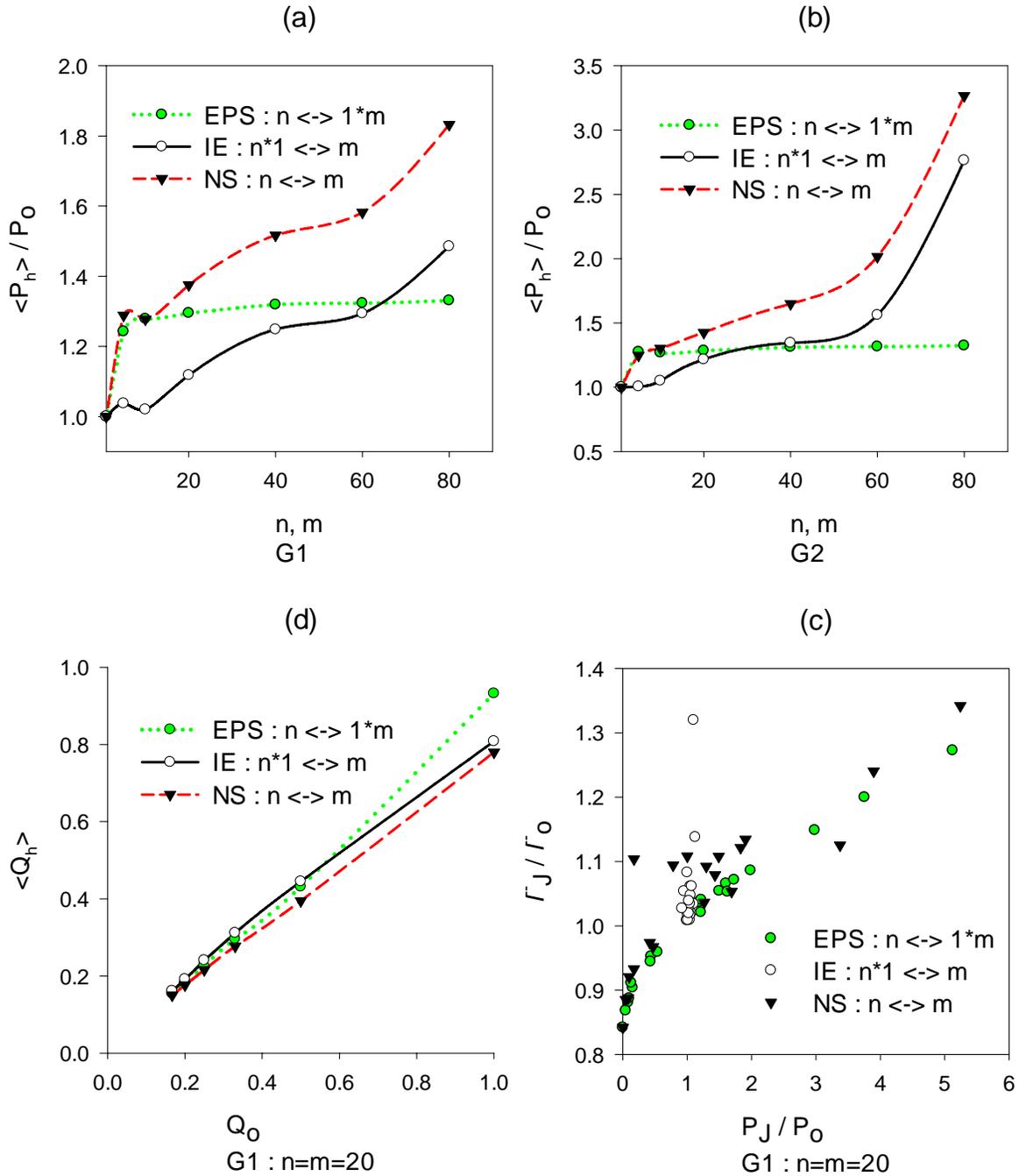

FIG. 2. (a) Ensemble averages of power enhancement in G1, (b) Ensemble averages of power enhancement in G2, (c) Eigenstates of decay rates and emitted power for one example case in the ensemble of *150* evaluations - *J* represents individual emitters in the case of IE evaluation, and the eigenstates in the other two evaluations, and (d) Ensemble averages of the normalized quantum efficiency of heterostructures as a function of quantum efficiency $Q_o$ of isolated emitters, for *n=m=20* in G1.

In the inverted complementary structure G2, the interaction between the emitters that are spread outside the cylinder is expected to be weaker. But the interaction among the metal particles is more significant here and the increase of LDOS has a larger effect on the power of emission. Thus the collective excitation regime of the emitters seems to have a smaller parametric range, and of lesser significance than in G1 as shown in Figure 2b. However, one should be cautioned that power enhancement is not a sufficient indicator of the underlying emission process. The lifetimes measured in such heterostructures are governed by the collective eigenstates; such as the example shown in Figure 2c. The above results in Figure 2a-2c were all evaluated for isolated emitters having a quantum efficiency of 1/6. However, Fig.2d shows that a change in the quantum efficiency of isolated emitters do not change the conclusions above; ensemble averages of *m=n=20* in G1 are evaluated over a range in $Q_o$ to show its almost linear effect on the normalized quantum efficiency of the heterostructure, $Q_h$ in Eq. (32). The apparent lifetimes measured in such heterostructures can be significantly dominated by the slower eigenstates of the collective excitation, and show notable differences with an isolated emitter interacting with the metal particles. Fig 3 shows emission dynamics for a few cases that use the calculated eigenstates and Eq. (34) to trace the decay of the collective excitations. Emitters of even moderate quantum efficiency ($Q_o$=0.5) show a clear shift in the decay curves of their collective emission. Their apparent decay seems to become slower with time; an effect of multiple eigenstates.

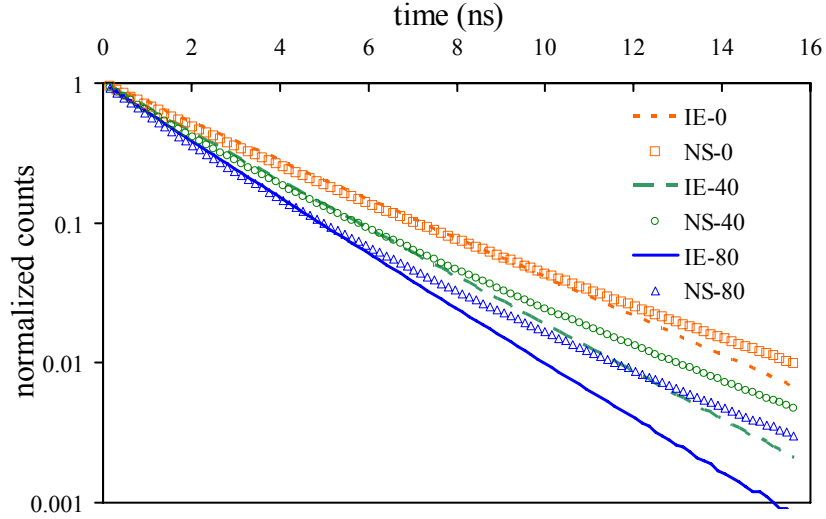

FIG. 3. Normalized time traces of decay representing 40 dipole emitters with varying number (0, 40, 80) of 3 nm radii gold particles in G1 heterostructure ensemble; $\hbar\omega_o = 2.21$ eV, $Q_o = 0.5$ and $\Gamma_o^r = 1/2 \times 10^9 \text{s}^{-1}$.

In all the above calculations, raw data of the ensemble show a pseudo-normal behavior confirming that the mean values represent the general behavior of a heterostructure (G1 or G2) for a specific case of *m, n*. The raw data of one such ensemble is shown in Fig. 4. As expected, IE and NS evaluations have marginally larger standard deviations in the ensembles due to their higher sensitivity to the permutation of metal particle locations. We have limited ourselves to results that highlight that collective emission can be notably different from independent emission. The full exploitation of the method presented here may need other numerical studies on a larger parametric space on many other heterostructures. These studies may help us to control exciton-plasmon couplings and resulting emission using low concentrations of even small metal particles; a regime that can certainly be very different from such effects on independent emitters.

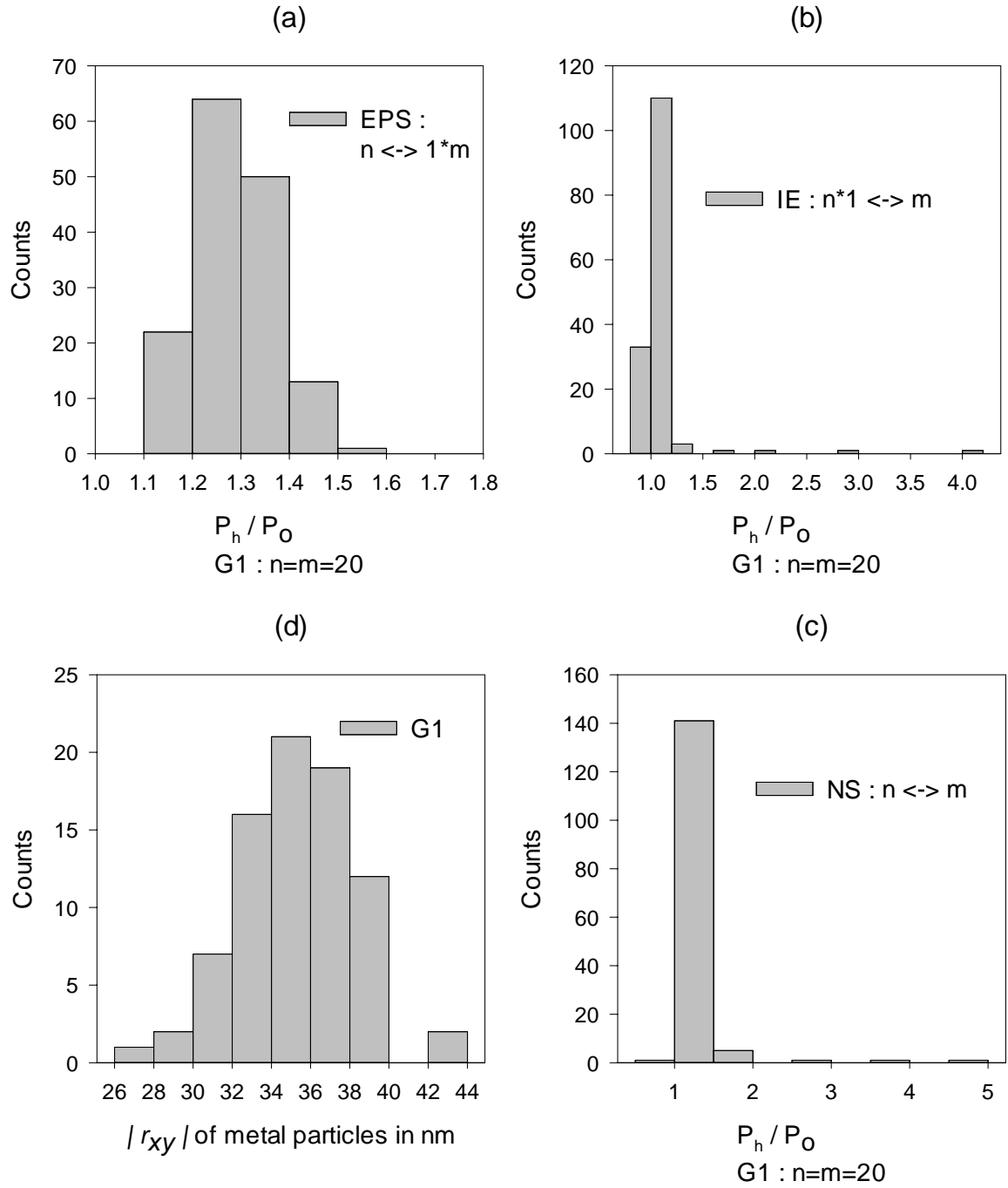

FIG. 4. Ensemble values of power enhancement in G1 for m=n=20: (a) EPS – Independent metal particles (b) IE – independent emitters, and (c) NS – interacting emitters and metal particles. (d) An example distribution of 80 metal particles in the XY plane of G1.

4. NOTE ON SURFACE INTERACTIONS:

The interaction of emitters and other bodies with a substrate/surface has been difficult to study because separable co-ordinates for analytical solutions do not exist, and Fresnel coefficients for near-field radiating sources do not have closed form relations.[35, 36] The determination of the reflected components of the surface on a radiating dipole in the near field involves computing the Sommerfeld relations; using numerical calculations of these integral relations, the field from a radiating dipole over a surface can be decomposed into cylindrical components parallel to the surface and a plane wave perpendicular to the surface. For any two dipoles $k$ and $j$ at any distance $z$ from the surface and distance $\rho$ from each other along the surface, the integral in Eq. (35) has to converge over the complex domain of wave vector magnitudes $k_\rho$ with appropriate branch cuts as the required reflection coefficient $R_s$ involves its quadratic roots.[37, 38] The reflected component of the electric field due to a radiating dipole in the near field of a surface can then be represented as in Eq. (36).[39, 40]

$$\frac{e^{ikr}}{4\pi r} = \frac{i}{4\pi}\int_0^\infty \frac{k_\rho}{k_z} J_o(\rho k_\rho) R_s \exp(-ik_z|z|) dk_\rho \tag{35}$$

$$\mathbf{E}_{surface,j} = \sum_{k=1}^{N}[\overline{S}_{jk} + \frac{k_s^2(k_o^2 - k_s^2)}{\varepsilon_0(k_o^2 + k_s^2)} \mathbf{G}_o^I(\mathbf{r}_j,\mathbf{r}_k;\omega)] \cdot \mathbf{p}_k \tag{36}$$

where $\overline{S}_{jk}$ is a 3 x 3 matrix containing Sommerfeld integral terms of the field for dipoles $k$ and $j$, and $k_o$ and $k_s$ are the wave numbers in free-space/homogeneous background and the surface respectively. For a surface with its normal in the $z$ direction, the image dyadic Green's function matrix is defined as

$$\mathbf{G}_o^I(\mathbf{r}_j,\mathbf{r}_k;\omega) = -\mathbf{G}_o(\mathbf{r}_j,\mathbf{r}_k;\omega) \cdot \mathbf{I}_R \text{ where } \mathbf{I}_R \text{ is the reflection dyad } \mathbf{I}_R = \mathbf{e}_x\mathbf{e}_x + \mathbf{e}_y\mathbf{e}_y - \mathbf{e}_z\mathbf{e}_z \tag{37}$$

The *surface-modified* Green dyads of the background medium $\mathbf{G}_o^s(\mathbf{r}_j,\mathbf{r}_k;\omega)$ can be introduced into Eq. (9) in place of $\mathbf{G}_o(\mathbf{r}_j,\mathbf{r}_k;\omega)$ for computing the self-energy matrices of a heterostructure on a surface, and they are

$$\mathbf{G}_o^s(\mathbf{r}_j,\mathbf{r}_k;\omega) = \overline{S}_{jk} + \mathbf{G}_o(\mathbf{r}_j,\mathbf{r}_k;\omega) \cdot (\mathbf{I} - \frac{k_s^2(k_o^2 - k_s^2)}{\varepsilon_0(k_o^2 + k_s^2)}\mathbf{I}_R) \qquad (38)$$

5. CONCLUSION:

A method to evaluate characteristics of eigenstates of an interacting set of *n* dipole emitters and other polarizable matter of generalized geometries was presented. The role of self-energy matrices in the estimation of their collective eigenstates was described, and the required relations for decomposition of their radiative and non-radiative parts were also derived. Closed form relations to evaluate Green tensors using the global matrices of Green dyads coupling entity pairs in free-space (or a homogeneous background medium) were produced. These relations were also shown to satisfy general laws of Physics such as a Lagrangian solution of the problem, and the optical theorem for many point emitters and other polarizable bodies interacting among themselves. The possible modifications for special limiting cases were discussed and a method to evaluate the collective interactions with a surface was presented. Moreover, the significant effects of *n* coupled emitters interacting with *m* polarizable bodies were highlighted using numerical results. The heterostructures used in these numerical examples provided a concise view of the rich behavior of emission possible in collective systems. This method of enumerating eigenstates of emission in strongly interacting emitter-matter systems provides a new path to deeper understanding of optical metamaterials.

ACKNOWLEDGEMENTS:

I am grateful to Bobby Philip, Oak Ridge National Laboratory, for many fruitful discussions and his comments on this manuscript.

APPENDIX: Pustovit-Shahbazyan model for m (independent) metal particles

The self-energy matrix of an ensemble of dipole emitters interacting with a single spherical metal particle was described by Pustovit and Shahbazyan, and an analytical solution of the green tensors was derived under long-wavelength approximation.[10] There, the phase of the oscillators was fixed while the amplitudes were modified due to the common field; as described by the self-energy matrix of the ensemble.

$$\Sigma_{jk}(\omega) = \frac{2\pi q^2 \omega_o}{mc^2} \mathbf{e}_j \cdot \mathbf{G}(\mathbf{r}_j, \mathbf{r}_k; \omega) \cdot \mathbf{e}_k$$

where $\Gamma_o^r = \frac{2kq^2\omega_o}{3mc^2}$ and $\omega \approx \omega_o$ (A1)

The modified Green dyads coupling the dipole emitters $P_j \mathbf{e}_j$ in the presence of a single spherical metal particle centered at $\mathbf{r}_p$ were derived. They satisfy the following relations where $\varepsilon_p, \varepsilon_o$ and $\theta$ are the permittivity of metal, free-space and the step function relating them, for a spherical particle of radius $R$.

$$\frac{4\pi\omega^2}{c^2} \sum_k \mathbf{G}_p(\mathbf{r}, \mathbf{r}_k; \omega) \cdot \mathbf{e}_k P_k e^{i\phi_k} = \mathbf{E}(\mathbf{r}, \omega) \text{ where}$$

$$k^2 \varepsilon(\mathbf{r}) \cdot \mathbf{G}_p(\mathbf{r}, \mathbf{r}_j; \omega) - \nabla \times \nabla \times \mathbf{G}_p(\mathbf{r}, \mathbf{r}_j; \omega) = \delta(\mathbf{r} - \mathbf{r}_j)$$ (A2)

and $\varepsilon(\mathbf{r}) = \varepsilon_p \theta(R - \|\mathbf{r} - \mathbf{r}_p\|) + \varepsilon_0 \theta(\|\mathbf{r} - \mathbf{r}_p\| - R)$

The Green dyadic between the dipole emitters at $\mathbf{r}_j, \mathbf{r}_k$ due to all the metal particles *not* interacting among each other, is then given by a sum of these modified Green dyads in the presence of all the *single* metal particles at the locations $\mathbf{r}_p$.

$$\mathbf{G}(\mathbf{r}_j,\mathbf{r}_k;\omega) = \mathbf{G}_o(\mathbf{r}_j,\mathbf{r}_k;\omega) + \sum_p \mathbf{G}_p(\mathbf{r}_j,\mathbf{r}_k;\omega) \text{ and} \tag{A3}$$

$$\Sigma_{jk} = \frac{3\Gamma_o^r}{4k^3} \{ \frac{\mathbf{e}_j \cdot \mathbf{e}_k}{r_{jk}^3} - \frac{3(\mathbf{e}_j \cdot \mathbf{r}_{jk})(\mathbf{e}_k \cdot \mathbf{r}_{jk})}{r_{jk}^5} - \sum_l \alpha_l T_{jk}^l \}$$
$$- \frac{i\Gamma_o^r}{2} \{ \mathbf{e}_j \cdot \mathbf{e}_k - \alpha_1 K_{jk}^1 + |\alpha_1|^2 T_{jk}^1 \} - \delta_{jk} \frac{i\Gamma_o^{nr}}{2} \tag{A4}$$

where

$$T_{jk}^l = \sum_p \frac{4\pi}{2l+1} \sum_{-l}^{+l} [\mathbf{e}_j \cdot \psi_{lm}(\mathbf{r}_j - \mathbf{r}_p)][\mathbf{e}_k \cdot \psi_{lm}(\mathbf{r}_k - \mathbf{r}_p)] \text{ and}$$
$$K_{jk}^l = \sum_p \frac{4\pi}{2l+1} \sum_{-l}^{+l} [\mathbf{e}_j \cdot \psi_{lm}(\mathbf{r}_j - \mathbf{r}_p)][\mathbf{e}_k \cdot \chi_{lm}(\mathbf{r}_k - \mathbf{r}_p)] \tag{A5}$$
$$\psi_{lm}(\mathbf{r}) = \nabla[\frac{Y_{lm}(\mathbf{r})}{r^{l+1}}], \chi_{lm}(\mathbf{r}) = \nabla[r^l Y_{lm}(\mathbf{r})]$$

$Y_{lm}(\mathbf{r})$ are the spherical harmonics and specifically, the dipole mode components are given by

$$K_{jk}^1 = \sum_p \frac{1}{\|\mathbf{r}_j - \mathbf{r}_p\|^3} \{ \mathbf{e}_j \cdot \mathbf{e}_k - \frac{3(\mathbf{e}_j \cdot (\mathbf{r}_j - \mathbf{r}_p))(\mathbf{e}_k \cdot (\mathbf{r}_j - \mathbf{r}_p))}{\|\mathbf{r}_j - \mathbf{r}_p\|^2} \}$$
$$T_{jk}^1 = \sum_p \frac{1}{\|\mathbf{r}_k - \mathbf{r}_p\|^3 \|\mathbf{r}_j - \mathbf{r}_p\|^3} \{ \mathbf{e}_j \cdot \mathbf{e}_k - ...$$
$$\frac{3(\mathbf{e}_j \cdot (\mathbf{r}_j - \mathbf{r}_p))(\mathbf{e}_k \cdot (\mathbf{r}_j - \mathbf{r}_p))}{\|\mathbf{r}_j - \mathbf{r}_p\|^2} - \frac{3(\mathbf{e}_j \cdot (\mathbf{r}_k - \mathbf{r}_p))(\mathbf{e}_k \cdot (\mathbf{r}_k - \mathbf{r}_p))}{\|\mathbf{r}_k - \mathbf{r}_p\|^2} + ... \tag{A6}$$
$$\frac{9(\mathbf{e}_k \cdot (\mathbf{r}_k - \mathbf{r}_p))(\mathbf{e}_j \cdot (\mathbf{r}_j - \mathbf{r}_p))((\mathbf{r}_k - \mathbf{r}_p) \cdot (\mathbf{r}_j - \mathbf{r}_p))}{\|\mathbf{r}_j - \mathbf{r}_p\|^2 \|\mathbf{r}_k - \mathbf{r}_p\|^2} \}$$

The self energy matrix of the ensemble of emitters can be decomposed into its radiative and non-radiative decay parts as

$$\Gamma^r_{jk} = \Gamma^r_o \{\mathbf{e}_j \cdot \mathbf{e}_k - \alpha'_1 K^1_{jk} + |\alpha_1|^2 T^1_{jk}\} \text{ and}$$

$$\Gamma^{nr}_{jk} = \frac{3\Gamma^r_o}{2k^3} \sum_l \alpha''_l T^l_{jk} - \delta_{jk} \frac{i\Gamma^{nr}_o}{2} \quad (A7)$$

where $\alpha_l = \alpha'_l + i\alpha''_l$

The *l*-pole polarizability of the spherical metal particle is given by

$$\alpha_l(\omega) = a^{2l+1} \left[ \frac{\varepsilon(\omega, \mathbf{r}_p) - \varepsilon}{\varepsilon(\omega, \mathbf{r}_p) + (1+1/l)\varepsilon} \right], \text{ where } a \text{ is radius of particle at } \mathbf{r}_p \quad (A8)$$